\documentclass{article} 

\usepackage[numbers]{natbib}       
\usepackage{float}                 

\usepackage{amsmath}              
\usepackage{algorithm}            
\usepackage{algorithmic}          

\usepackage{multirow}
\usepackage{booktabs}             
\usepackage{nicefrac}             
\usepackage{microtype}            
\usepackage{lipsum}               

\usepackage[utf8]{inputenc}       
\usepackage[T1]{fontenc}          
\usepackage{amsfonts}             

\usepackage{hyperref}             
\usepackage{url}                  

\usepackage{graphicx}
\graphicspath{{./images/}}        

\title{\textnormal{ReCoSeg++}: Extended Residual-Guided Cross-Modal Diffusion for Brain Tumor Segmentation}

\author{
Sara Yavari\textsuperscript{1} \quad 
Rahul Nitin Pandya\textsuperscript{1} \quad 
Jacob Furst\textsuperscript{1} \\
\textsuperscript{1}School of Computing, DePaul University, Chicago, IL, USA \\
\texttt{\{syavari, npandya, jfurst\}@depaul.edu}
}

\begin{document}

\maketitle 

\begin{abstract}
Accurate segmentation of brain tumors in MRI scans is critical for clinical diagnosis and treatment planning.  We propose a semi-supervised, two-stage framework that extends the ReCoSeg approach to the larger and more heterogeneous BraTS 2021 dataset, while eliminating the need for ground-truth masks for the segmentation objective.  In the first stage, a residual-guided denoising diffusion probabilistic model (DDPM) performs cross-modal synthesis by reconstructing the T1ce modality from FLAIR, T1, and T2 scans.  The residual maps, capturing differences between predicted and actual T1ce images, serve as spatial priors to enhance downstream segmentation. In the second stage, a lightweight U-Net takes as input the concatenation of residual maps, computed as the difference between real T1ce and synthesized T1ce, with T1, T2, and FLAIR modalities to improve whole tumor segmentation.  To address the increased scale and variability of BraTS 2021, we apply slice-level filtering to exclude non-informative samples and optimize thresholding strategies to balance precision and recall. Our method achieves a Dice score of $93.02\%$ and an IoU of $86.7\%$ for whole tumor segmentation on the BraTS 2021 dataset, outperforming the  ReCoSeg baseline on BraTS 2020 (Dice: $91.7\%$, IoU: $85.3\%$), and demonstrating improved accuracy and scalability for real-world, multi-center MRI datasets.

\textbf{keywords}: Brain Tumor Segmentation, Diffusion Models, Residual Maps, Semi-Supervised Learning. 
\end{abstract}
\section{Introduction}
\label{sec:intro}
Brain tumor segmentation from MRI scans is a critical task in neuro-oncology, facilitating accurate diagnosis, informed treatment planning, and long-term patient monitoring \cite{Bauer2013Survey}. Deep learning models particularly U-Net \cite{re1} and its 3D extension, 3D U-Net \cite{re2} have become standard approaches for this application, achieving high segmentation accuracy through end-to-end supervised learning frameworks. However, these models typically require large volumes of annotated data and involve high computational costs \cite{re3}, which limits their scalability in real-world clinical environments where annotations may be scarce and imaging modalities inconsistently available. To address these challenges, recent research has turned to generative modeling techniques. Notably, Denoising Diffusion Probabilistic Models (DDPMs) have been employed for cross-modal synthesis to reconstruct missing MRI modalities \cite{re4}. For example, DDMM Synth~\cite{DDMM} uses diffusion models to accurately generate the T1ce modality from other available sequences. Recently, DDPMS Known for their ability to generate high-quality, diverse images \cite{Ho2020}, beyond image generation, they have shown strong performance in a range of vision tasks, DPMs have also demonstrated strong performance in various tasks such as image editing \cite{re7}, super-resolution \cite{re8}, and segmentation \cite{re9}, underscoring their flexibility and robustness. This shows that diffusion models are versatile and reliable, working well across many different computer vision tasks. Prominent large-scale diffusion models-such as DALL·E 2 \cite{re10}, Imagen \cite{re11}, and Stable Diffusion \cite{re12} have shown remarkable capabilities in high-quality image generation \cite{re13}; \cite{re14}.

At the core of diffusion models is a two-stage process structured as a Markov chain. In the forward stage, Gaussian noise is progressively added to the input image over multiple timesteps, eventually transforming it into pure noise. The reverse process then learns to gradually denoise this input, reconstructing the original image step-by-step \cite{Ho2020}. This iterative denoising framework underpins the impressive performance of diffusion models in visual generation tasks. Originally, diffusion models were applied to domains where absolute ground truth data is unavailable or ill-defined. Motivated by the strengths of DPMs, this paper focuses on improving binary brain tumor segmentation in MRI using a semi-supervised approach that combines cross-modal synthesis with residual-based attention, following the approach described in previous work \cite{re15}.

\begin{figure}[H]
\centering
\includegraphics[width=0.9\linewidth]{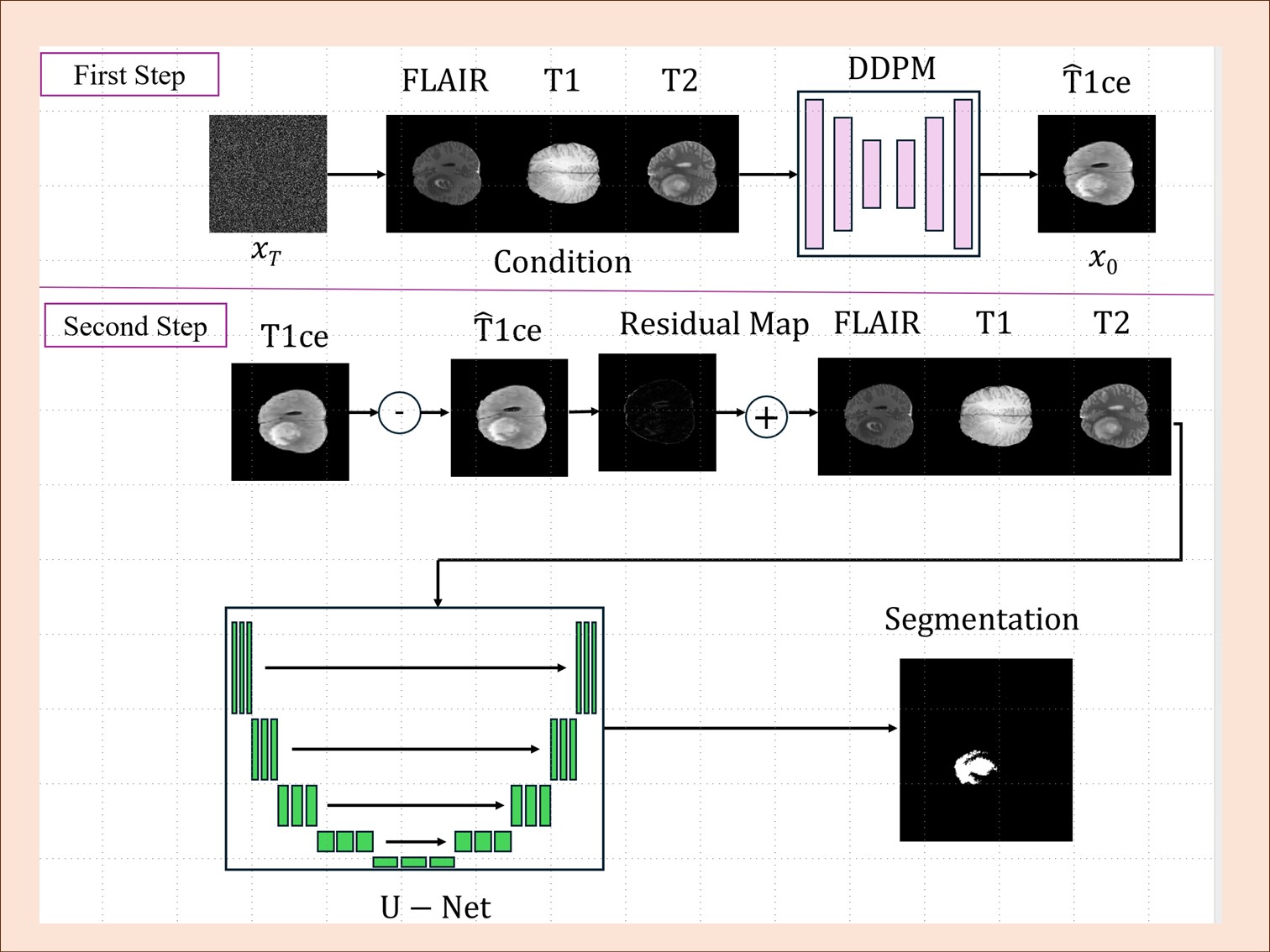}
\caption{
Overview of the proposed ReCoSeg framework for brain tumor segmentation using multi-modal MRI. The method consists of two key stages: (1) A cross-modal synthesis stage, where a Denoising Diffusion Probabilistic Model (DDPM) is employed to generate the T1ce modality from the available FLAIR, T1, and T2 sequences. This addresses the challenge of missing modalities and enhances the completeness of input data; (2) A residual-guided segmentation stage, in which a residual map is computed by subtracting the synthesized T1ce from the real T1ce. This residual map highlights regions of structural discrepancy, often indicative of tumor tissue, and is concatenated with the original modalities. The combined inputs are then fed into a lightweight 2D U-Net, where the residuals act as attention cues to guide the segmentation process. This two-phase design improves boundary localization, reduces reliance on dense annotations, and enhances both interpretability and efficiency in clinical workflows.
}
\label{fig:Recog}
\end{figure}

The proposed ReCoSeg++ framework adopts a two-step strategy to enhance brain tumor segmentation from MRI sequences, following the approach described in previous work \cite{shortmidl}. In the first step, a DDPM synthesizes the T1ce modality from the available FLAIR, T1, and T2 scans to address the challenge of missing modalities and provide a more complete set of inputs for downstream analysis. The synthesized T1ce is then compared to its real counterpart to compute a residual map that highlights regions with structural discrepancies, often corresponding to tumor tissue. In the second step, this residual map is concatenated with the original MRI modalities and input into a lightweight 2D U-Net for segmentation. This residual-guided attention mechanism enables the model to focus on informative regions, improving segmentation accuracy while reducing reliance on densely annotated data, as illustrated in Figure~\ref{fig:Recog}.

\section{Related work}
\label{sec:rel}
Magnetic Resonance Imaging (MRI) is an essential tool in modern radiology, especially for neurological evaluations, due to its exceptional soft tissue contrast and ability to acquire images across multiple planes and sequences \cite{reLi}. Unlike Computed Tomography (CT), which relies on ionizing X-rays, MRI uses strong magnetic fields and radio frequency (RF) pulses to generate high-resolution cross-sectional images of brain anatomy \cite{reMRI}. A key advantage of MRI is its support for multi-sequence imaging-such as T1-weighted (T1w), T2-weighted (T2w), Fluid-Attenuated Inversion Recovery (FLAIR), and Diffusion-Weighted Imaging (DWI)-each providing unique, complementary insights critical for tumor detection and characterization \cite{reAu}.

The emergence of deep learning has significantly advanced medical image segmentation, particularly for MRI-based brain tumor analysis. Convolutional Neural Networks (CNNs) have driven this progress \cite{Md,re121}, with architectures like U-Net and its variants effectively modeling spatial hierarchies for accurate tumor localization \cite{Sun}. Models such as TransBTS \cite{Wang}, which integrates attention mechanisms, and CANet \cite{Liu}, which captures context-aware features, have improved performance, especially near tumor boundaries. Nonetheless, challenges remain in segmenting small or irregular tumors due to loss of spatial detail and limited feature expressiveness \cite{reAu}. To overcome these issues, extensions like U-Net++ \cite{Zhou} with nested skip connections, 3D U-Net \cite{re28}, and V-Net \cite{Mil} for volumetric imaging have been developed. Attention U-Net \cite{ok}, TransUNet \cite{Chen}, and Swin-UNet \cite{Cao} leverage attention and transformer mechanisms to capture both local details and global context essential for delineating complex boundaries. Despite these innovations, the encoder-decoder paradigm remains central to state-of-the-art models, as exemplified by nnU-Net \cite{Is} and its improvements. For instance, \cite{Luu} introduced axial attention into the decoder, \cite{Fut} optimized foreground voxel representation with increased network depth, and \cite{Sid} enhanced robustness through adaptive ensembling under data perturbations, underscoring the ongoing evolution of CNN-based approaches in brain tumor segmentation.

Several studies have shown the effectiveness of integrating VAEs and other generative models into segmentation architectures \cite{gms,generative}. For example, \cite{re14} introduced a 3D brain tumor segmentation model with a VAE branch to regularize the encoder-decoder framework, improving generalization and performance. Similarly, \cite{re15} proposed a two-stage cascaded approach combining VAEs with attention gates, achieving high accuracy across distinct tumor sub-regions by capturing both global context and local details. Expanding unsupervised feature learning, \cite{re18} developed a dual autoencoder model with Singular Value Decomposition (SVD) for optimized feature extraction and dimensionality reduction, enhancing segmentation of complex tumor structures. The Dual Residual Multi-Variational Autoencoder (DRM-VAE) introduced by \cite{re19} uses multiple VAEs and residual connections to handle missing MRI modalities and maintain robust performance in incomplete data scenarios. Beyond VAEs, Generative Adversarial Networks (GANs), introduced by \cite{re20}, have also proven useful in medical imaging. GANs employ adversarial training between a generator and discriminator to synthesize realistic MRI images, which can augment training datasets, address annotation scarcity, and improve segmentation accuracy \cite{re21, re22}. Advanced GAN-based models like Vox2Vox further improve 3D brain tumor segmentation by capturing spatial dependencies and producing detailed segmentation maps with high Dice scores \cite{re23}.

Diffusion models have recently emerged as a powerful approach for brain tumor segmentation in MRI scans \cite{re24,re25}. Their inherent stochasticity introduces controlled randomness during training, enhancing model flexibility and enabling better adaptation to complex tumor morphologies and structural variations \cite{re25}. In this context, \cite{re26} applied Denoising Diffusion Probabilistic Models (DDPMs) for brain MRI segmentation, offering a novel approach that synthesizes labeled data to reduce dependence on manual annotations. Although effective, this method involves substantial computational costs and long inference times. To address this, \cite{re27} proposed the PD-DDPM model, a diffusion-based framework that iteratively refines segmentation predictions through reverse denoising. By leveraging pre-segmentation outputs and predicting noise based on forward diffusion dynamics, PD-DDPM reduces computational time while maintaining segmentation accuracy and structural fidelity.

MedSegDiff \cite{re31} advanced diffusion probabilistic models (DPMs) for medical image segmentation by introducing a dynamic conditional encoding strategy and a Frequency-Filter (FF) resolver to mitigate high-frequency noise. Building on this, MedSegDiff-V2 \cite{re32} incorporated a transformer-based architecture with Gaussian spatial attention blocks to better align noise patterns with semantic features, improving noise estimation and segmentation accuracy. In parallel, SegDiff \cite{re30} proposed a framework that integrates image data directly into the reverse diffusion process, enabling iterative refinement of segmentation outputs at each denoising step for greater accuracy and structural consistency. Complementing these efforts, DMCIE \cite{DMCIE2025} introduced a hybrid method that leverages a base segmentation model alongside a diffusion process guided by the discrepancy between predicted and ground truth masks, focusing refinement on uncertain regions to enhance segmentation performance and efficiency.

In contrast to prior works, our proposed framework introduces a segmentation-aware diffusion approach that explicitly reconstructs residual maps by comparing synthesized and real T1ce modalities without using ground-truth masks. These residuals serve as attention cues, highlighting tumor regions and concatenating with multi-modal MRI inputs to guide a lightweight U-Net for accurate segmentation. This design improves tumor boundary localization while maintaining modularity, interpretability, and robustness in cases of missing or corrupted modalities.
\section{Methodology}
This section consists of two subsections. In Subsection~\ref{subsec:diffusion}, a diffusion model is employed to synthesize the T1ce modality from the available FLAIR, T1, and T2 MRI scans. The difference between the synthesized and real T1ce images, captured as residuals, highlights potential tumor regions. These residuals are then utilized in Subsection~\ref{subsubsec:unet} as attention cues within a lightweight U-Net architecture for segmentation. This approach reduces the dependence on densely annotated labels while improving segmentation performance.

\subsection{T1ce Synthesis Using Diffusion Models}
\label{subsec:diffusion}
In the first stage of our proposed ReCoSeg++ framework, we employ an enhanced Conditional Denoising Diffusion Probabilistic Model (DDPM) to synthesize the T1ce modality from the available FLAIR, T1, and T2 MRI sequences. The DDPM follows a forward-reverse diffusion process \cite{Ho2020}, which gradually adds Gaussian noise to an input image over \( T \) steps and then learns to reverse this noising process to recover the original image.

Let \( \mathbf{x}_0 \sim q(\mathbf{x}_0) \) denote a clean image sampled from the real distribution of the T1ce modality. The forward diffusion process adds noise at each step \( t \in \{1, \ldots, T\} \) as:
\begin{equation}
q(\mathbf{x}_t | \mathbf{x}_{t-1}) = \mathcal{N}(\mathbf{x}_t; \sqrt{1 - \beta_t} \mathbf{x}_{t-1}, \beta_t \mathbf{I}),
\end{equation}
where \( \beta_t \) is a variance schedule typically increasing with \( t \).

By recursively applying this, we can sample \( \mathbf{x}_t \) directly from \( \mathbf{x}_0 \) via:
\begin{equation}
q(\mathbf{x}_t | \mathbf{x}_0) = \mathcal{N}(\mathbf{x}_t; \sqrt{\bar{\alpha}_t} \mathbf{x}_0, (1 - \bar{\alpha}_t) \mathbf{I}),
\end{equation}
with \( \bar{\alpha}_t = \prod_{s=1}^{t} (1 - \beta_s) \).

The reverse process is modeled using a neural network \( \epsilon_\theta \) to predict the noise:
\begin{equation}
p_\theta(\mathbf{x}_{t-1} | \mathbf{x}_t) = \mathcal{N}(\mathbf{x}_{t-1}; \mu_\theta(\mathbf{x}_t, t), \sigma_t^2 \mathbf{I}),
\end{equation}
where \( \mu_\theta \) is computed using the noise prediction \( \epsilon_\theta(\mathbf{x}_t, t) \).

The training objective is to minimize the simplified loss:
\begin{equation}
\mathcal{L}_{\text{simple}} = \mathbb{E}_{\mathbf{x}_0, \epsilon, t} \left[ \left\| \epsilon - \epsilon_\theta(\sqrt{\bar{\alpha}_t} \mathbf{x}_0 + \sqrt{1 - \bar{\alpha}_t} \epsilon, t) \right\|_2^2 \right].
\end{equation}

In our cross-modal setup, the conditioning input \( \mathbf{c} \in \mathbb{R}^{H \times W \times 3} \) includes the FLAIR, T1, and T2 images. We concatenate these channels and use channel-wise conditioning with attention blocks to capture multi-modal context. The network is trained to generate \( \mathbf{x}_0^{\text{T1ce}} \) from noise, conditioned on \( \mathbf{c} \). This is implemented via conditional DDPM:
\begin{equation}
\epsilon_\theta = \epsilon_\theta(\mathbf{x}_t, t \mid \mathbf{c}),
\end{equation}
which learns to map noisy latent inputs to clean T1ce outputs guided by the input modalities.

Self-attention layers are integrated to better capture long-range anatomical dependencies in 3D volumes. We further improve stability and sharpness of generated T1ce images through an optimized noise variance schedule with better balancing across timesteps. Additionally, we used a plateau-based learning rate scheduler that reduces the learning rate from 3e-4 to 1.5e-4 when the model stops improving, helping it train more steadily.

To further encourage semantic accuracy, we jointly optimize a reconstruction loss combining Binary Cross-Entropy (BCE) and Dice loss between the generated \( \hat{\mathbf{x}}_0^{\text{T1ce}} \) and the ground truth \( \mathbf{x}_0^{\text{T1ce}} \):
\begin{equation}
\mathcal{L}_{\text{recon}} = \lambda_1 \cdot \mathcal{L}_{\text{BCE}} + \lambda_2 \cdot \mathcal{L}_{\text{Dice}},
\end{equation}
where \( \lambda_1, \lambda_2 \) are hyperparameters balancing the contributions of the two losses.

Once synthesis is complete, we compute the residual map \( \mathbf{R} \) as the absolute pixel-wise difference between the generated and real T1ce scans:
\begin{equation}
\mathbf{R} = |\hat{\mathbf{x}}_0^{\text{T1ce}} - \mathbf{x}_0^{\text{T1ce}}|.
\label{eq:residual}
\end{equation}
Unlike static error maps, these residuals act as soft attention priors that highlight reconstruction uncertainty and anomaly cues, guiding the segmentation network to focus on clinically relevant regions with higher precision.
\subsection{Residual-Aware Tumor Segmentation via Lightweight U-Net}
\label{subsubsec:unet} 
In the second stage of the ReCoSeg++ framework, the focus shifts from synthesis to segmentation. After generating the synthetic T1ce modality via the diffusion model, we compute a residual map \( \mathbf{R} \), defined in Equation~\ref{eq:residual}, capturing voxel-wise differences between the generated and real T1ce images. These residuals act as dynamic, error-aware soft attention priors, highlighting regions of high reconstruction uncertainty that are likely to contain tumor tissue.

To improve robustness, we introduce a simple threshold calibration step on the residual map, adjusting its dynamic range to emphasize informative error regions while suppressing noise from well-reconstructed background. To ensure consistent intensity scaling, residuals are min-max normalized. This calibrated residual map is then concatenated with the original input modalities (FLAIR, T1, and T2) to form the final segmentation input:
\begin{equation}
\mathbf{X}_{\text{seg}} = \text{Concat}(\mathbf{x}^{\text{FLAIR}}, \mathbf{x}^{\text{T1}}, \mathbf{x}^{\text{T2}}, \mathbf{R}),
\end{equation}
resulting in a four-channel input that combines anatomical information with error-aware residual guidance.

This multi-channel input is processed by a lightweight 2D U-Net \cite{liao}, which maintains high segmentation accuracy while being computationally efficient for deployment in resource-constrained environments. The error-aware residual priors help focus the model's attention on ambiguous regions where tumor boundaries are uncertain. The segmentation network produces a binary tumor mask \( \hat{\mathbf{y}} \in [0,1]^{H \times W} \), where each value represents the probability of tumor presence. The model is trained using a compound loss function that combines Binary Cross-Entropy (BCE) and Dice loss:

\begin{equation}
\mathcal{L}_{\text{seg}} = \lambda_1 \cdot \mathcal{L}_{\text{BCE}}(\hat{\mathbf{y}}, \mathbf{y}) + \lambda_2 \cdot \mathcal{L}_{\text{Dice}}(\hat{\mathbf{y}}, \mathbf{y}),
\end{equation}

where \( \mathbf{y} \) is the ground truth segmentation mask, and \( \lambda_1, \lambda_2 \) are hyperparameters controlling the balance between losses.
By integrating calibrated residual maps as soft attention cues, the U-Net is guided to focus more precisely on suspicious regions. This approach improves tumor localization performance and reduces the dependency on extensive labeled data, resulting in a clinically interpretable and efficient segmentation pipeline.

\section{Experimental Setup}
\subsection{Dataset}
We evaluate our method using both the BraTS 2020 and BraTS 2021 datasets, which provide multi-center, multimodal 3D MRI scans with expert-annotated tumor segmentation masks. The BraTS2020 dataset includes 355 subjects, while BraTS2021 expands this to over 400 subjects, introducing more heterogeneous patient cases and improved annotation quality. Each scan includes four MRI modalities: FLAIR, T1, T2, and T1ce. Following standard practice, we use T2, FLAIR, and T1 as conditioning inputs for the synthesis of the T1ce modality.

To ensure consistent anatomical context, all volumes undergo pre-processing steps to harmonize resolution and spatial coverage. we discard the top 26 and bottom 80 axial slices in each volume, where tumor presence is rare, retaining 78 informative slices per subject. Each slice is normalized by clipping the top and bottom 1st percentile of intensities and applying z-score normalization. The volumes are then center-cropped or resized to a consistent shape of (C, D, H, W) = (3, 78, 120, 120), where the three channels correspond to FLAIR, T1, and T2 modalities.

Additionally, we apply slice-level filtering to exclude slices without any tumor signal in the T1ce ground truth. This reduces background bias and ensures that the model focuses on anatomically relevant regions with potential pathology, reducing class imbalance and avoiding overfitting. This targeted sampling significantly lowers computational overhead, enabling efficient training and experimentation while preserving performance and generalization.

All scans are resampled to a consistent voxel spacing of 1 mm isotropic resolution to standardize spatial detail across subjects. We also preserve the real distribution variability of tumor location, size, and appearance across institutions and scanners to improve the robustness of the model and generalization to unseen clinical settings.

\subsection {Implementation Details}
All models in our study, were implemented in PyTorch 2.0 and trained on an NVIDIA RTX 3090 GPU with 24\,GB of memory. Given GPU limits, batch sizes and architecture complexity varied. Accordingly, batch sizes and optimization settings were carefully tuned. For 2D-based models, we extract axial slices and resize them to 120×120 pixels. Voxel intensities are clipped to the 1st-99th percentile, z-score normalized, and processed with slice-level filtering to exclude non-informative slices lacking tumor regions. To make the model more robust, we apply random image augmentations using Albumentations, including flips, rotations, affine transforms, and brightness/contrast changes, with a probability of 0.3 to 0.5 during training.

In contrast, UNet3D processes volumetric patches across the axial dimension and requires significantly more memory per sample. To accommodate this, we reduce the batch size and apply gradient accumulation when necessary. Despite its higher computational cost, UNet3D serves as an important baseline for comparing performance with context-rich volumetric segmentation approaches. Batch sizes were set to 4 for the 2D diffusion model and 2 for 3D segmentation models, using PyTorch DataLoader with shuffling and worker settings adapted to CPU availability. This strategy ensures batches contain diverse anatomical regions and subjects to avoid overfitting.

The diffusion model in ReCoSeg++ and DDMM-Synth is a conditional denoising diffusion probabilistic model (DDPM), trained for 1000 timesteps using the Adam optimizer. We use a plateau-based learning rate scheduler that reduces the learning rate from 3e-4 to 1.5e-4 upon stagnation, stabilizing training and improving convergence. Training was stabilized using early stopping based on the validation Dice score to prevent overfitting during prolonged training. The conditioning inputs are FLAIR, T1, and T2, while the target is T1ce. Both ReCoSeg++ and DDMM-Synth share identical backbones and training hyperparameters to ensure a fair comparison.

For segmentation, all models-including the ReCoSeg++ lightweight U-Net-are trained using a hybrid loss combining Binary Cross-Entropy (BCE) and Dice loss. This formulation ensures voxel-level accuracy while maintaining strong spatial overlap with ground truth tumor regions. We also incorporate threshold calibration, empirically selecting $\tau=0.3$ to binarize the sigmoid outputs, based on both visual assessment and validation performance (e.g., Dice and IoU). The threshold $\tau$ was empirically determined through a validation sweep across $\tau \in \{0.3, 0.4, 0.5\}$, with $\tau = 0.3$ yielding the best Dice-IoU tradeoff. The entire training pipeline maintains consistent random seed initialization, data split strategy, and preprocessing to ensure that observed performance differences can be attributed to the models' architectures themselves rather than implementation variance. The large VRAM capacity of the RTX 3090 GPU enables training of both 2D and 3D models within reasonable memory and time budgets.

\subsection {Evaluation Metrics}
To assess the performance of our proposed ReCoSeg++ framework and its baselines, we adopt two widely used evaluation metrics in medical image segmentation, Dice Similarity Coefficient (Dice) \cite{sta} and Intersection over Union (IoU) \cite{eel}. These metrics quantify the spatial overlap between predicted and ground truth tumor masks and are particularly informative for evaluating models in imbalanced binary segmentation tasks such as brain tumor delineation.

All models are trained using a hybrid loss composed of Binary Cross-Entropy (BCE) and Dice loss, which balances voxel-wise classification with region-based segmentation performance.  All segmentation networks target the Whole Tumor (WT) class label.

Table~\ref{tab:results} reports the segmentation results across the models. Among all models, ReCoSeg++ achieves the highest performance, with a Dice score of 0.917 and an IoU of 0.853. Compared to the baseline DDMM-Synth \cite{DDMM}, which also leverages a diffusion model for T1ce synthesis, ReCoSeg++ demonstrates a clear improvement, validating the effectiveness of residual-guided segmentation.

The volumetric UNet3D \cite{re28} baseline performs better than the fully supervised UNet2D \cite{ron}, highlighting the benefit of 3D context for tumor segmentation. However, it remains inferior to DDMM-Synth  \cite{DDMM} and ReCoSeg \cite{shortmidl}. DDMM-Synth \cite{DDMM} provides a richer segmentation but is surpassed by Recoseg++ with superior localization and enhanced residual-guided attention. In particular, ReCoSeg++ outperforms all baselines while maintaining a lightweight and modular architecture, making it well-suited for real-world clinical deployment where computational resources are limited, especially when ground truth is scarce or unavailable.

\begin{table}[t]
\small
\setlength{\tabcolsep}{5pt}
\renewcommand{\arraystretch}{1.1}
\centering
\caption{Comparison with baselines on BraTS2020 and BraTS2021 validation sets. All scores are reported as percentages.}
\label{tab:results}
\begin{tabular}{lcccc}
\hline
Model & \multicolumn{2}{c}{BraTS2020} & \multicolumn{2}{c}{BraTS2021} \\
      & Dice (\%) & IoU (\%) & Dice (\%) & IoU (\%) \\
\hline
UNet2D~\cite{ron}               & 78.4 & 73.6 & 87.3  & 81.0 \\
UNet3D~\cite{re28}                & 84.2 & 74.3 & 88.1  & 81.7 \\
DDMM-Synth~\cite{DDMM}          & 87.2 & 81.1 & 90.9  & 85.1 \\
ReCoSeg ~\cite{shortmidl}      & 91.7   & 85.3   & 91.2  & 83.6 \\
ReCoSeg++ (Ours)                  & 89.8 & 86.4 & 93.02 & 86.7 \\
\hline
\end{tabular}
\end{table}

As shown in Table~\ref{tab:results}, the integration of cross-modal synthesis, residual-guided attention, and uncertainty-aware error maps allows ReCoSeg++ to achieve accurate segmentation even without access to ground-truth T1ce at inference time. The model also remains lightweight and suitable for deployment in low-resource clinical settings.

\subsection {Results}
ReCoSeg++ represents a significant advancement over the previous ReCoSeg framework~\cite{shortmidl} by introducing multiple architectural, training methods, and data handling, which together lead to higher segmentation accuracy and better generalization. A key difference lies in the dataset used for training and evaluation. While the ReCoSeg\cite{shortmidl} was developed on BraTS 2020 with approximately 369 subjects, ReCoSeg++ leverages the larger and more diverse BraTS 2021 dataset with over 1,250 subjects. This richer dataset enables better learning of varied tumor morphologies, consistent annotations, and improved generalization to rare or atypical tumor presentations.

As illustrated in Figure~\ref{fig:Recog}, ReCoSeg++ enhances Stage 1 by adopting an improved DDPM for T1ce synthesis. This version integrates a hybrid BCE + Dice loss, optimized noise schedules, and channel-wise conditioning, resulting in sharper predicted T1ce maps with better preservation of tumor contrast. The residual maps-computed as voxel-wise absolute differences between ground truth and predicted T1ce-serve as dynamic error-aware priors that guide Stage 2, segmentation more precisely toward tumor regions.

In Stage 2, ReCoSeg++ employs a lightweight but still effective 2D U-Net segmentation head designed for axial slices, which fuses FLAIR, T1, T2, and residual maps into a four-channel input. Unlike the static thresholding used in the original ReCoSeg~\cite{shortmidl}, ReCoSeg++ systematically calibrates segmentation thresholds ($\tau = 0.3, 0.4, 0.5$) via visual and quantitative analysis, selecting $\tau=0.3$ as the optimal operating point balancing precision and recall.

\begin{figure}[!htbp]
\centering
\includegraphics[width=1.2\textwidth]{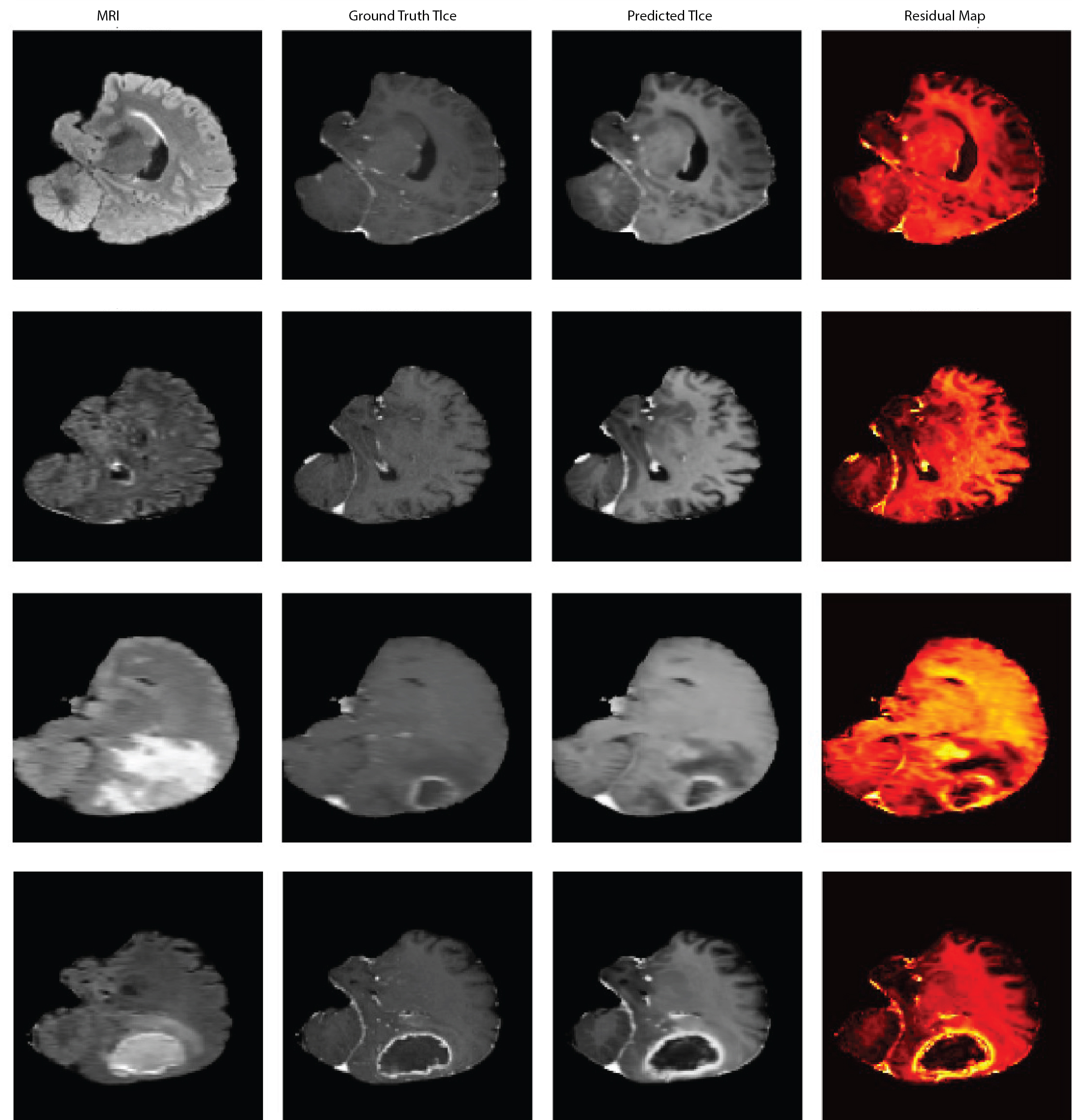} 
\caption{Qualitative results of ReCoSeg++ for T1ce reconstruction: from left to right – multi-modal MRI input, ground truth T1ce, predicted T1ce, and residual map highlighting reconstruction errors.}
\label{fig:recoseg_visuals}
\end{figure}

\begin{figure}[H]
\centering
\includegraphics[width=1.2\linewidth]{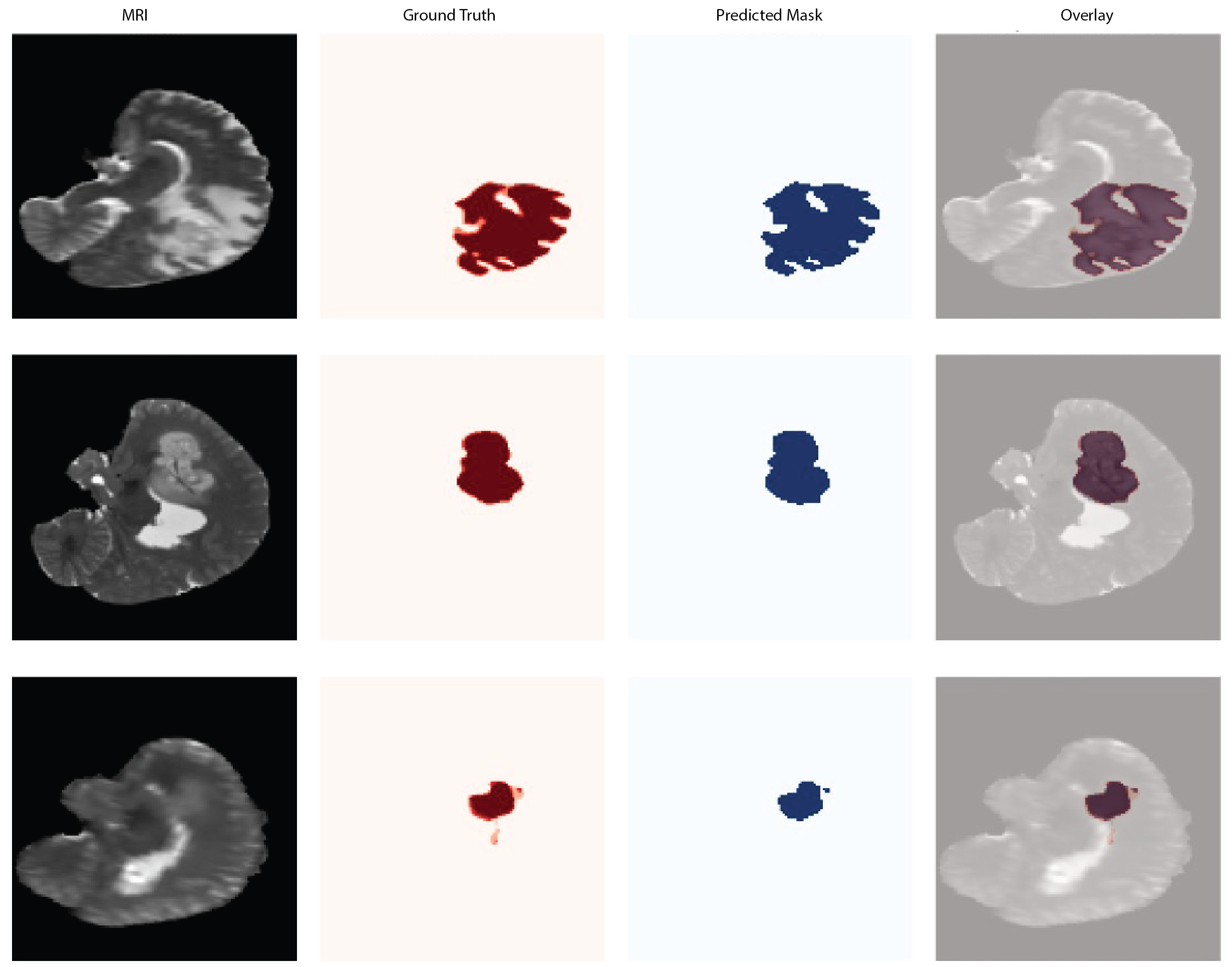}
\caption{Qualitative segmentation results of ReCoSeg++ on the BraTS2020 dataset. From left to right: input MRI slice, ground truth tumor mask, predicted tumor mask, and overlay of the prediction on the MRI. The model accurately captures tumor boundaries with strong spatial agreement.}
\label{fig:brats}
\end{figure}

\begin{figure}
\centering
\includegraphics[width=1\linewidth]{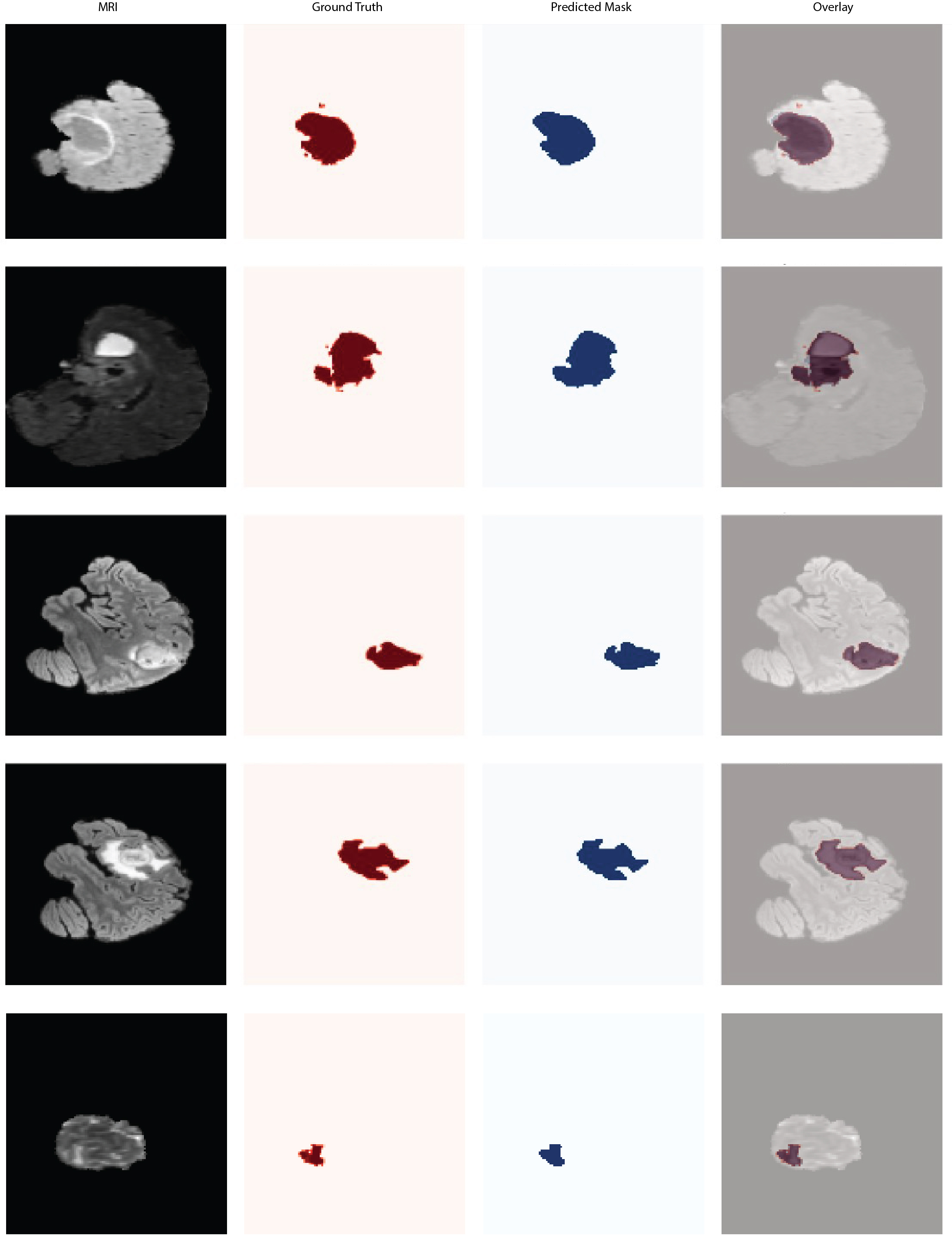}
\caption{Qualitative segmentation results of ReCoSeg++ on the BraTS2021 dataset. From left to right: input MRI slice, ground truth tumor mask, predicted tumor mask, and overlay. Despite increased anatomical variability, the model demonstrates robust generalization and precise tumor localization.}
\label{fig:sg}
\end{figure}

Quantitative results demonstrate the superiority of ReCoSeg++, achieving a Dice score of approximately 93.02\% and an IoU of 86.7\%, surpassing both its predecessor and multiple baseline models. This improvement reflects not only better data and architecture but also the benefit of residual-guided, attention-enhancing error maps that focus learning on clinically relevant tumor features. Furthermore, ReCoSeg++ maintains a modular, computationally efficient design suitable for deployment in resource-constrained clinical settings. Qualitative results, shown in Figure~\ref{fig:recoseg_visuals}, demonstrate the model’s capacity to synthesize the contrast-enhanced T1ce modality from multi-modal MRI inputs. The residual maps highlight areas of disagreement between the predicted and ground truth T1ce images, particularly along tumor boundaries, suggesting that the model effectively captures tumor-specific features and discrepancies useful for downstream segmentation tasks. Qualitative segmentation results are shown in Figure~\ref{fig:brats} and Figure~\ref{fig:sg}, highlighting ReCoSeg++'s performance on BraTS2020 and BraTS2021, respectively. Each figure presents a visual comparison of the input MRI, ground truth tumor mask, predicted mask, and an overlay of the prediction. As seen in Figure~\ref{fig:brats}, the model captures tumor boundaries with high spatial precision and minimal over-segmentation on BraTS2020, demonstrating strong alignment with expert annotations. Figure~\ref{fig:sg} further illustrates the model’s robustness on the more diverse BraTS2021 dataset, where tumor appearance and size vary significantly. Even in challenging cases, ReCoSeg++ produces accurate and consistent segmentation masks, validating its generalization capability across datasets.

\section{Ablation Study}
We corroborate the effectiveness of our method by ablating individual design choices in the ReCoSeg++ framework, including the residual generation strategy, threshold calibration for segmentation binarization, and segmentation backbone efficiency. The residual maps are critical for guiding the network toward uncertain and tumor-prone regions. Two variants were compared: static residuals computed as simple absolute difference maps without diffusion modeling, and dynamic residuals generated via the DDPM-based synthesis with improved noise schedules and channel-wise conditioning. The dynamic residuals demonstrated superior localization of enhancing tumor boundaries, resulting in an approximate +1.2\% mean Dice score improvement on the validation set.

A systematic threshold calibration was performed by evaluating segmentation masks at $\tau = 0.3, 0.4, 0.5$. Visual and quantitative analyses confirmed that $\tau = 0.3$ provided the best balance of precision and recall, especially in ambiguous tumor boundaries. Higher thresholds (e.g., $\tau = 0.5$) led to more conservative masks with reduced recall, while $\tau = 0.3$ maintained sensitivity with manageable false positives. This calibration step delivered an approximate +0.8\% improvement in Dice score compared to a fixed default threshold. Finally, to ensure feasibility for deployment in resource-constrained clinical settings, a lightweight 2D U-Net\cite{liao} segmentation head was adopted instead of a standard 3D U-Net. Despite its reduced complexity, this design maintained competitive accuracy. While 3D U-Net baselines offered strong context modeling with Dice scores around 88\%, the proposed 2D U-Net with residual guidance achieved higher accuracy (approximately 93\% Dice), underscoring the benefits of dynamic residual guidance, careful calibration, and efficient design for clinical applications.

\begin{figure}
\centering
\includegraphics[width=0.95\linewidth]{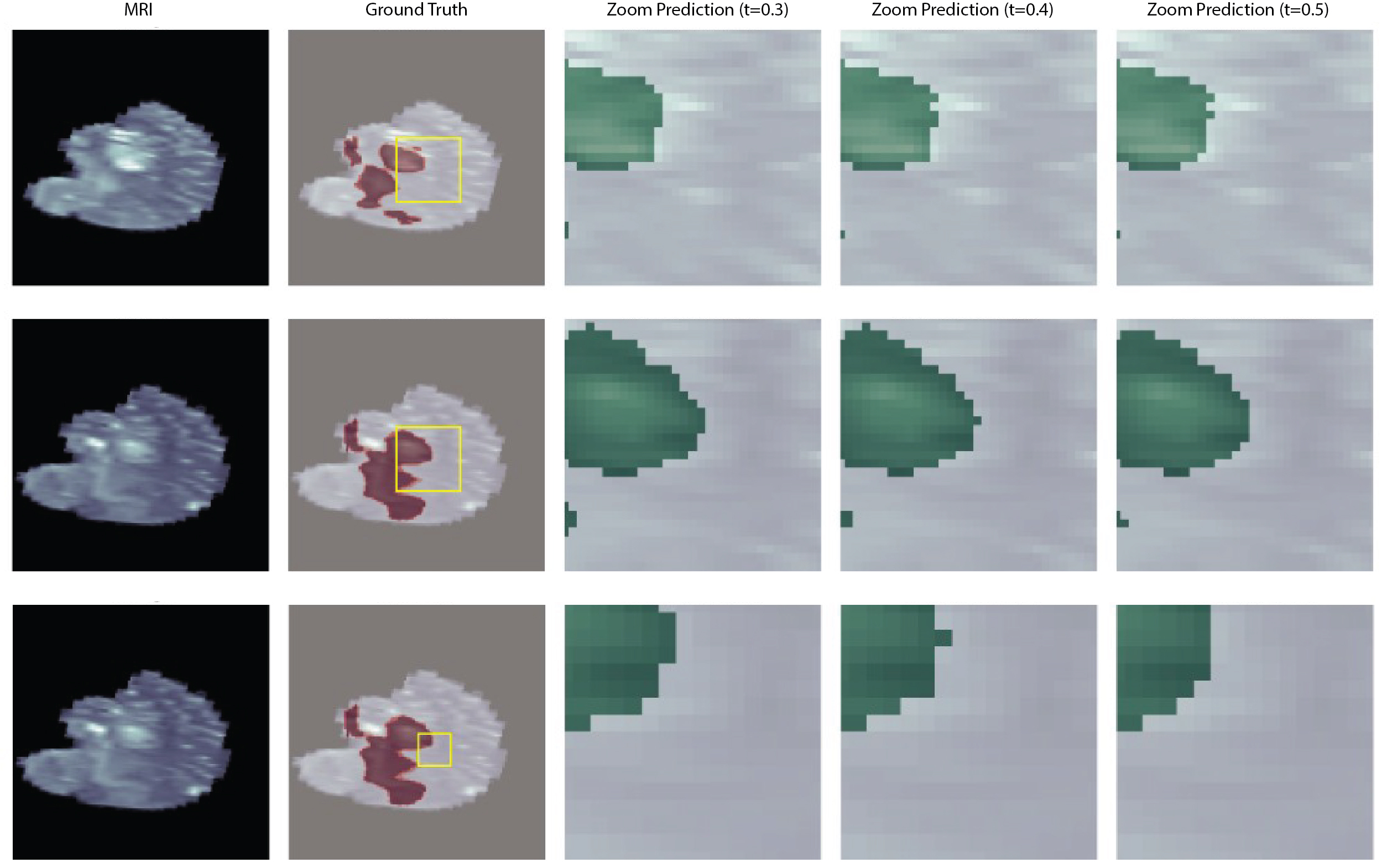}
\caption{Visualization of segmentation predictions under different binarization thresholds ($\tau=0.3$, $0.4$, and $0.5$). From left to right: input MRI, overlay with ground truth, and zoomed-in predictions at each threshold. Lower thresholds (e.g., $\tau=0.3$) result in better tumor boundary recall, while higher thresholds yield more conservative masks.}
\label{fig:threshold_zoom}
\end{figure}

Figure~\ref{fig:threshold_zoom} illustrates the effect of segmentation threshold calibration. By comparing predictions at $\tau=0.3$, $0.4$, and $0.5$, we observe that $\tau=0.3$ achieves the best balance between precision and recall, especially along tumor boundaries. Higher thresholds result in conservative predictions, potentially missing small or ambiguous regions.

\section{Conclusion}
We presented ReCoSeg++, an enhanced two-stage framework for brain tumor segmentation that integrates diffusion-based cross-modal synthesis with residual-guided attention. Using the differences between synthesized and real T1ce modalities, ReCoSeg++ generates dynamic residual maps that effectively localize tumor regions and guide the segmentation process. Extensive experiments on the BraTS2021 dataset demonstrate the effectiveness of our approach, achieving a Dice score of 93.02 and an IoU of 86.7 outperforming established baselines. Qualitative results further validate the model’s ability to delineate complex tumor boundaries, even in the absence of the T1ce modality during inference. Ablation studies confirm that residual guidance, threshold calibration, and the use of a lightweight 2D U-Net each contribute to the model’s strong performance. With its modular design and low computational cost, ReCoSeg++ offers a practical and accurate solution for real-world clinical deployment in brain tumor segmentation.

\bibliographystyle{unsrt}  

\bibliography{Ref}     

\begin{thebibliography}{10}

\bibitem{Bauer2013Survey}
Stefan Bauer, Roland Wiest, Lutz-P. Nolte, and Mauricio Reyes.
\newblock A survey of mri-based medical image analysis for brain tumor studies.
\newblock {\em Physics in Medicine and Biology}, 58(13):R97--R129, 2013.

\bibitem{re1}
Chenyang Liu, Rui Zhao, and Zhenwei Shi.
\newblock Remote-sensing image captioning based on multilayer aggregated transformer.
\newblock {\em IEEE Geoscience and Remote Sensing Letters}, 19:1--5, 2022.

\bibitem{re2}
Wenxuan Wang, Chen Chen, Meng Ding, Hong Yu, Sen Zha, and Jiangyun Li.
\newblock Transbts: Multimodal brain tumor segmentation using transformer.
\newblock In {\em International Conference on Medical Image Computing and Computer-Assisted Intervention (MICCAI)}, pages 109--119. Springer, 2021.

\bibitem{re3}
Chenyang Liu, Rui Zhao, and Zhenwei Shi.
\newblock Remote-sensing image captioning based on multilayer aggregated transformer.
\newblock {\em IEEE Geoscience and Remote Sensing Letters}, 19:1--5, 2022.

\bibitem{re4}
Wenxuan Wang, Chen Chen, Meng Ding, Hong Yu, Sen Zha, and Jiangyun Li.
\newblock Transbts: Multimodal brain tumor segmentation using transformer.
\newblock In {\em International Conference on Medical Image Computing and Computer-Assisted Intervention (MICCAI)}, pages 109--119. Springer, 2021.

\bibitem{DDMM}
Yixiao Wang, Wenjia Zhang, Hanyu Li, Yule Qin, Ji~Wu, and Yongchao Xu.
\newblock Ddmm-synth: A denoising diffusion model for cross-modal medical image synthesis.
\newblock {\em arXiv preprint arXiv:2303.15770}, 2023.

\bibitem{Ho2020}
J.~Ho, A.~Jain, and P.~Abbeel.
\newblock Denoising diffusion probabilistic models.
\newblock In {\em Advances in Neural Information Processing Systems}, volume~33, pages 6840--6851. Curran Associates, Inc., 2020.

\bibitem{re7}
B.~Kawar et~al.
\newblock Imagic: Text-based real image editing with diffusion models.
\newblock In {\em Proceedings of the IEEE/CVF Conference on Computer Vision and Pattern Recognition (CVPR)}, pages 6007--6017. IEEE, 2023.

\bibitem{re8}
S.~Gao et~al.
\newblock Implicit diffusion models for continuous super-resolution.
\newblock In {\em Proceedings of the IEEE/CVF Conference on Computer Vision and Pattern Recognition (CVPR)}, pages 10021--10030. IEEE, 2023.

\bibitem{re9}
M.~Sun, W.~Huang, and Y.~Zheng.
\newblock Instance-aware diffusion model for gland segmentation in colon histology images.
\newblock In {\em Proceedings of the International Conference on Medical Image Computing and Computer-Assisted Intervention (MICCAI)}, pages 662--672. Springer, 2023.

\bibitem{re10}
Aditya Ramesh, Prafulla Dhariwal, Alex Nichol, Casey Chu, and Mark Chen.
\newblock Hierarchical text-conditional image generation with clip latents.
\newblock {\em arXiv preprint arXiv:2204.06125}, 2022.

\bibitem{re11}
Chitwan Saharia, William Chan, Saurabh Saxena, Lala Li, Jay Whang, Emily Denton, Seyed Kamyar~Seyed Ghasemipour, Burcu~Karagol Ayan, S.~Sara Mahdavi, Rapha~Gontijo Lopes, et~al.
\newblock Photorealistic text-to-image diffusion models with deep language understanding.
\newblock {\em arXiv preprint arXiv:2205.11487}, 2022.

\bibitem{re12}
Robin Rombach, Andreas Blattmann, Dominik Lorenz, Patrick Esser, and Björn Ommer.
\newblock High-resolution image synthesis with latent diffusion models.
\newblock In {\em Proceedings of the IEEE/CVF Conference on Computer Vision and Pattern Recognition (CVPR)}, pages 10684--10695. IEEE, 2022.

\bibitem{re13}
Rui Zhao and Zhenwei Shi.
\newblock Text-to-remote-sensing-image generation with structured generative adversarial networks.
\newblock {\em IEEE Geoscience and Remote Sensing Letters}, 19:1--5, 2021.

\bibitem{re14}
Ian Goodfellow, Jean Pouget-Abadie, Mehdi Mirza, Bing Xu, David Warde-Farley, Sherjil Ozair, Aaron Courville, and Yoshua Bengio.
\newblock Generative adversarial networks.
\newblock {\em Communications of the ACM}, 63(11):139--144, 2020.

\bibitem{re15}
Chenggang Lyu and Hai Shu.
\newblock A two-stage cascade model with variational autoencoders and attention gates for mri brain tumor segmentation.
\newblock {\em Brainlesion}, pages 435--447, 2020.

\bibitem{shortmidl}
Sara Yavari, Rahul Pandya, and Jacob Furst.
\newblock Recoseg: Residual guided cross-modal diffusion for efficient brain tumor segmentation.
\newblock In {\em Proceedings of the Medical Imaging with Deep Learning (MIDL)}. DePaul University, Chicago, IL, USA, MIDL, 2025.
\newblock Accepted for publication.

\bibitem{reLi}
Wenqing Li, Wenhui Huang, and Yuanjie Zheng.
\newblock Corrdiff: Corrective diffusion model for accurate mri brain tumor segmentation.
\newblock {\em IEEE Journal of Biomedical and Health Informatics}, 28(3):1587, 2024.

\bibitem{reMRI}
Aparna Pai, Rohil Shetty, Brendan Hodis, and Yuvraj~S. Chowdhury.
\newblock Magnetic resonance imaging physics, 2023.
\newblock StatPearls [Internet]. Treasure Island (FL): StatPearls Publishing; Last updated: April 2, 2023.

\bibitem{reAu}
Philipp Kickingereder, Fabian Isensee, Irada Tursunova, et~al.
\newblock Automated quantitative tumour response assessment of mri in neuro-oncology with artificial neural networks: A multicentre, retrospective study.
\newblock {\em The Lancet Oncology}, 2019.

\bibitem{Md}
Md~Kamrul~Hasan Khan, Wenjing Guo, Jie Liu, Fan Dong, Zoe Li, Tucker~A. Patterson, and Huixiao Hong.
\newblock Machine learning and deep learning for brain tumor mri image segmentation.
\newblock {\em Experimental Biology and Medicine (Maywood)}, 248(21):1974--1992, 2023.

\bibitem{re121}
Asma Alshuhail, Arastu Thakur, R.~Chandramma, T.~R. Mahesh, Ahlam Almusharraf, V.~Vinoth Kumar, and Surbhi~Bhatia Khan.
\newblock Refining neural network algorithms for accurate brain tumor classification in mri imagery.
\newblock {\em BMC Medical Imaging}, 24:118, 2024.

\bibitem{Sun}
Ming Sun, Weilin Huang, and Yefeng Zheng.
\newblock Instance-aware diffusion model for gland segmentation in colon histology images.
\newblock In {\em Proceedings of the International Conference on Medical Image Computing and Computer-Assisted Intervention (MICCAI)}, pages 662--672. Springer, 2023.

\bibitem{Wang}
Wenqi Wang, Chen Chen, Mingyu Ding, Jingyun Li, Hongming Yu, and Shi Zha.
\newblock Transbts: Multimodal brain tumor segmentation using transformer.
\newblock In {\em Proceedings of the International Conference on Medical Image Computing and Computer-Assisted Intervention (MICCAI)}, pages 109--119. Springer, 2021.

\bibitem{Liu}
Zongwei Liu, Guotai Wang, Lequan Yu, Yizhe Zhang, Xiaomeng Li, and Pheng-Ann Heng.
\newblock Canet: Context aware network for brain glioma segmentation.
\newblock {\em IEEE Transactions on Medical Imaging}, 40(7):1763--1777, Jul 2021.

\bibitem{Zhou}
Zongwei Zhou, Md~Mahfuzur~Rahman Siddiquee, Nima Tajbakhsh, and Jianming Liang.
\newblock Unet++: A nested u-net architecture for medical image segmentation.
\newblock {\em CoRR}, abs/1807.10165, 2018.

\bibitem{re28}
Özgün Çiçek, Ahmed Abdulkadir, Soeren~S. Lienkamp, Thomas Brox, and Olaf Ronneberger.
\newblock 3d u-net: Learning dense volumetric segmentation from sparse annotation.
\newblock In {\em International Conference on Medical Image Computing and Computer-Assisted Intervention (MICCAI)}, pages 424--432. Springer, 2016.

\bibitem{Mil}
Fausto Milletari, Nassir Navab, and Seyed-Ahmad Ahmadi.
\newblock V-net: Fully convolutional neural networks for volumetric medical image segmentation.
\newblock In {\em Proceedings of the 2016 Fourth International Conference on 3D Vision (3DV)}, pages 565--571. IEEE, 2016.

\bibitem{ok}
Ozan Oktay, Jo~Schlemper, Loic~Le Folgoc, Matthew Lee, Mattias Heinrich, Kazunari Misawa, Kensaku Mori, Steven McDonagh, Nils~Y. Hammerla, Bernhard Kainz, Ben Glocker, and Daniel Rueckert.
\newblock Attention u-net: Learning where to look for the pancreas.
\newblock {\em CoRR}, abs/1804.03999, 2018.

\bibitem{Chen}
Jieneng Chen, Yongyi Lu, Qihang Yu, Xiangde Luo, Ehsan Adeli, Yan Wang, Le~Lu, Alan~L. Yuille, and Yuyin Zhou.
\newblock Transunet: Transformers make strong encoders for medical image segmentation.
\newblock {\em CoRR}, abs/2102.04306, 2021.

\bibitem{Cao}
Huwaisong Cao, Yueyue Wang, Joy~Iong Zong, Jianfei Yang, Ying Chen, Zhen Li, Maoqing Tian, and Li~Zhang.
\newblock Swin-unet: Unet-like pure transformer for medical image segmentation.
\newblock {\em CoRR}, abs/2105.05537, 2022.

\bibitem{Is}
Fabian Isensee, Jens Petersen, Andre Klein, David Zimmerer, Paul~F. Jaeger, Simon Kohl, Jakob Wasserthal, Gregor Köhler, Tobias Norajitra, Sebastian Wirkert, and Klaus~H. Maier-Hein.
\newblock nnu-net: A self-configuring method for deep learning-based biomedical image segmentation.
\newblock {\em Nature Methods}, 18:203--211, 2021.

\bibitem{Luu}
Huan~Minh Luu and Sung-Hong Park.
\newblock Extending nnu-net for brain tumor segmentation.
\newblock In {\em International MICCAI Brainlesion Workshop}, volume 12963 of {\em Lecture Notes in Computer Science}, pages 173--186. Springer, 2021.

\bibitem{Fut}
Michał Futrega, Alexandre Milesi, Michał Marcinkiewicz, and Pablo Ribalta.
\newblock Optimized u-net for brain tumor segmentation.
\newblock In {\em International MICCAI Brainlesion Workshop}, volume 12963 of {\em Lecture Notes in Computer Science}, pages 15--29. Springer, 2021.

\bibitem{Sid}
Md~Mahfuzur~Rahman Siddiquee and Andriy Myronenko.
\newblock Redundancy reduction in semantic segmentation of 3d brain tumor mris.
\newblock {\em arXiv preprint}, arXiv:2111.00742, 2021.

\bibitem{gms}
Jiayu Huo, Xi~Ouyang, Sébastien Ourselin, and Rachel Sparks.
\newblock Generative medical segmentation, 2024.
\newblock arXiv preprint arXiv:2403.17876.

\bibitem{generative}
Li~Zhang, Basu Jindal, Ahmed Alaa, Robert Weinreb, David Wilson, Eran Segal, James Zou, and Pengtao Xie.
\newblock Generative ai enables medical image segmentation in ultra low-data regimes, 2024.
\newblock arXiv preprint arXiv:2408.17421.

\bibitem{re18}
K.~Aswani and D.~Menaka.
\newblock A dual autoencoder and singular value decomposition based feature optimization for the segmentation of brain tumor from mri images.
\newblock {\em BMC Medical Imaging}, 21:82, 2021.

\bibitem{re19}
Yian Zhu, Shaoyu Wang, Yun Hu, and Xiao Ma.
\newblock Drm-vae: A dual residual multi variational auto-encoder for brain tumor segmentation with missing modalities.
\newblock In {\em 2021 IEEE 4th International Conference on Electronics and Communication Engineering (ICECE)}. IEEE, 2021.

\bibitem{re20}
Ian Goodfellow, Jean Pouget-Abadie, Mehdi Mirza, Bing Xu, David Warde-Farley, Sherjil Ozair, Aaron Courville, and Yoshua Bengio.
\newblock Generative adversarial networks.
\newblock {\em arXiv preprint}, 2014.

\bibitem{re21}
H.~C. Shin, N.~A. Tenenholtz, J.~K. Rogers, C.~G. Schwarz, M.~L. Senjem, J.~L. Gunter, K.~P. Andriole, and M.~Michalski.
\newblock Medical image synthesis for data augmentation and anonymization using generative adversarial networks.
\newblock In {\em Lecture Notes in Computer Science}, volume 11037, pages 1--11. Springer, 2018.

\bibitem{re22}
Xin Yi, Ekta Walia, and Paul Babyn.
\newblock Generative adversarial network in medical imaging: A review.
\newblock {\em Medical Image Analysis}, 2019.

\bibitem{re23}
Marco~Domenico Cirillo, David Abramian, and Anders Eklund.
\newblock Vox2vox: 3d-gan for brain tumour segmentation.
\newblock {\em arXiv preprint}, 2020.

\bibitem{re24}
Yiqing Shen, Guannan He, and Mathias Unberath.
\newblock Promptable counterfactual diffusion model for unified brain tumor segmentation and generation with mris.
\newblock {\em arXiv preprint}, 2024.

\bibitem{re25}
Haokai Zhao, Haowei Lou, Lina Yao, Wei Peng, Ehsan Adeli, Kilian~M. Pohl, and Yu~Zhang.
\newblock Diffusion models for computational neuroimaging: A survey.
\newblock {\em arXiv preprint}, 2024.

\bibitem{re26}
Jonas Wolleb, Robin Sandkühler, Florian Bieder, Pietro Valmaggia, and Pascal~C. Cattin.
\newblock Diffusion models for implicit image segmentation ensembles.
\newblock In {\em International Conference on Medical Imaging with Deep Learning (MIDL)}. PMLR, 2022.

\bibitem{re27}
Xutao Guo, Yanwu Yang, Chenfei Ye, Shang Lu, Yang Xiang, and Ting Ma.
\newblock Accelerating diffusion models via pre-segmentation diffusion sampling for medical image segmentation.
\newblock {\em arXiv preprint}, 2022.

\bibitem{re31}
Jiachen Wu, Hao Fang, Yuying Zhang, Yubo Yang, and Yufeng Xu.
\newblock Medsegdiff: Medical image segmentation with diffusion probabilistic model.
\newblock {\em arXiv preprint}, 2022.

\bibitem{re32}
Junde Wu, Wei Ji, Rao Fu, Huazhu Fu, Min Xu, Yueming Jin, and Yanwu Xu.
\newblock Medsegdiff-v2: Diffusion based medical image segmentation with transformer.
\newblock {\em arXiv preprint}, 2023.

\bibitem{re30}
Tomer Amit, Eliya Nachmani, Tom Shaharbany, and Lior Wolf.
\newblock Segdiff: Image segmentation with diffusion probabilistic models.
\newblock {\em arXiv preprint}, 2021.

\bibitem{DMCIE2025}
Sara Yavari, Rahul~Nitin Pandya, and Jacob Furst.
\newblock Dmcie: Diffusion model with concatenation of inputs and errors to improve the accuracy of the segmentation of brain tumors in mri images.
\newblock {\em arXiv preprint arXiv:2507.00983}, Jul 2025.

\bibitem{liao}
Weibin Liao, Yinghao Zhu, Xinyuan Wang, Chengwei Pan, Yasha Wang, and Liantao Ma.
\newblock Lightm-unet: Mamba assists in lightweight unet for medical image segmentation.
\newblock {\em arXiv preprint arXiv:2403.05246}, 2024.

\bibitem{sta}
Kelly~H. Zou and et~al.
\newblock Statistical validation of image segmentation quality based on a spatial overlap index.
\newblock {\em Academic Radiology}, 11(2):178--189, 2004.

\bibitem{eel}
Tom Eelbode and et~al.
\newblock Optimization for medical image segmentation: Theory and practice when evaluating with dice score or jaccard index.
\newblock {\em IEEE Transactions on Medical Imaging}, 39(11):3679--3690, 2020.

\bibitem{ron}
Olaf Ronneberger, Philipp Fischer, and Thomas Brox.
\newblock U-net: Convolutional networks for biomedical image segmentation.
\newblock In {\em International Conference on Medical Image Computing and Computer-Assisted Intervention}, pages 234--241. Springer, 2015.

\end{thebibliography}
\end{document}